%% file: main.tex
\documentclass[sigconf,screen]{acmart}
\AtBeginDocument{%
  \providecommand\BibTeX{{%
    \normalfont B\kern-0.5em{\scshape i\kern-0.25em b}\kern-0.8em\TeX}}}

\usepackage{graphicx}
\usepackage{enumitem}
\usepackage{MnSymbol}
\usepackage{wasysym}
\usepackage{pifont}
\usepackage{multirow}
\usepackage{caption}
\usepackage{subcaption}

\iftrue
\usepackage{titlesec}
\titlespacing\section{0pt}{3pt plus 1pt minus 1pt}{0pt plus 1pt minus 1pt}
\titlespacing\subsection{0pt}{3pt plus 1pt minus 1pt}{0pt plus 1pt minus 1pt}
\titlespacing\subsubsection{0pt}{3pt plus 1pt minus 1pt}{2pt plus 1pt minus 1pt}
\fi

\setcopyright{acmcopyright}
\copyrightyear{2022}
\acmYear{2022}
\setcopyright{acmcopyright}\acmConference[DAC '22]{Proceedings of the 59th ACM/IEEE Design Automation Conference (DAC)}{July 10--14, 2022}{San Francisco, CA, USA}
\acmBooktitle{Proceedings of the 59th ACM/IEEE Design Automation Conference (DAC) (DAC '22), July 10--14, 2022, San Francisco, CA, USA}
\acmPrice{15.00}
\acmDOI{10.1145/3489517.3530452}
\acmISBN{978-1-4503-9142-9/22/07}

\begin{document}

\title{
\texttt{Eventor}: An Efficient \underline{Event}-Based Monocular Multi-View Stereo Accelera\underline{tor} on FPGA Platform
}
\titlenote{This work was supported by the National Natural Science Foundation of China (Grant No. 62072019). \\
Corresponding author: \textit{Jianlei Yang}, Email: \url{jianlei@buaa.edu.cn}}

\author{Mingjun Li$^1$, \quad Jianlei Yang$^1$, \quad Yingjie Qi$^1$, \quad Meng Dong$^2$, \quad Yuhao Yang$^1$,}
\author{Runze Liu$^3$, \quad Weitao Pan$^2$, \quad Bei Yu$^4$, \quad Weisheng Zhao$^1$}

\affiliation{
    \institution{
        $^1$Beihang University, Beijing, China \hspace{3em}
        $^2$Xidian University, Xi'an, Shaanxi, China \\
        $^3$Beijing Real Imaging Medical Technology Co., Ltd. \hspace{3em} 
        $^4$The Chinese University of Hong Kong, Hong Kong
    }
}


\input{0-abstract.tex}

\setcopyright{none}

\maketitle
\pagestyle{plain} 
\pagenumbering{gobble}

\input{1-introduction.tex}
\input{2-system.tex}
\input{3-architecture.tex}
\input{4-experiment.tex}
\input{5-conclusions.tex}

\bibliographystyle{unsrt}
\bibliography{ref}

\end{document}

%% file: 0-abstract.tex
\begin{abstract}

Event cameras are bio-inspired vision sensors that asynchronously represent pixel-level brightness changes as event streams.
Event-based monocular multi-view stereo (EMVS) is a technique that exploits the event streams to estimate semi-dense 3D structure with known trajectory.
It is a critical task for event-based monocular SLAM.
However, the required intensive computation workloads make it challenging for real-time deployment on embedded platforms.
In this paper, \texttt{Eventor} is proposed as a fast and efficient EMVS accelerator by realizing the most critical and time-consuming stages including event back-projection and volumetric ray-counting on FPGA.
Highly paralleled and fully pipelined processing elements are specially designed via FPGA and integrated with the embedded ARM as a heterogeneous system to improve the throughput and reduce the memory footprint.
Meanwhile, the EMVS algorithm is reformulated to a more hardware-friendly manner by rescheduling, approximate computing and hybrid data quantization.
Evaluation results on DAVIS dataset show that \texttt{Eventor} achieves up to $24\times$ improvement in energy efficiency compared with Intel i5 CPU platform.

\end{abstract}

%% file: 1-introduction.tex
\section{Introduction}
\label{sec:intro}

Event cameras are bio-inspired vision sensors developed in recent years \cite{brandli2014240}.
Different from traditional frame-based cameras which capture a scene as a synchronous sequence of 2D images, event cameras asynchronously measure brightness changes on each pixel and output \emph{event} streams.
An event encodes the timestamp, pixel coordinates and polarity of brightness changes.
Compared with traditional cameras, event cameras have numerous advantages: extremely high event rate ($>10^6$ events per second, events/s) and dynamic range (up to $130$ dB) while traditional cameras usually obtain $\sim 30$ FPS and $65$ dB, respectively \cite{son20174}.
Additionally, event cameras only require a very low data rate (KB vs. MB) by removing an amount of the inherent redundancy of standard cameras, thus making it quite efficient.

The unique properties of event cameras make them as ideal sensors for running visual SLAM systems on low-power embedded platforms, such as robots and drones, for real-time applications.
The event-based monocular visual SLAM systems involve event-based 3D reconstruction which aims to estimate the depth information and the structure of the scene from event cameras.
Unlike the multi-view stereo methods, the monocular methods only require a single event camera which do not pursue instantaneous depth estimation, but rather depth estimation for SLAM \cite{cadena2016past}.
Recently, the event-based monocular multi-view stereo (EMVS) technique has received particular attention, since its performance will greatly affect the overall performance of visual SLAM systems \cite{zhou2021event}\cite{zhou2018semi}.
However, it is very challenging to unlock the benefits of event cameras for monocular multi-view stereo applications on embedded platforms for real-time purpose.
This is due to the fact that event cameras represent a paradigm shift in acquisition of visual information, thus requiring novel algorithms and specified hardware design \cite{gallego2020event}. 
Previous accelerators designed for traditional intensity-frame-based multi-view stereo algorithms cannot be directly applied for the event-based algorithms.

Several previous algorithms have been proposed for EMVS implementations \cite{rebecq2018emvs}\cite{kim2016real}\cite{gallego2018unifying} but all of them could only run on relatively powerful CPU or GPU platforms.
Aiming to improve the computational efficiency of EMVS, an event-based space-sweep method \cite{rebecq2018emvs} is proposed by back-projecting events to create a ray density volume \cite{collins1996space}, and then find local maxima of ray density to estimate the scene structure.
Such an efficient EMVS implementation integrated with an event-based visual odometry (EVO) system \cite{rebecq2016evo} could process $1.2$ million events/s when running with a single core of Intel x86 CPU, and $4.7$ million events/s with $4$ cores \cite{rebecq2018emvs}.
However, running the EMVS algorithms on multi-core x86 CPUs is not practical for embedded EVO applications.
Another event processing pipeline is proposed in \cite{kim2016real} by utilizing three filters running in parallel to jointly estimate the motion of the event camera and 3D map.
Such an approach only runs on GPUs for real-time performance and cannot process high event rate input (up to $1$M events/s).
A unified event processing framework is proposed in \cite{gallego2018unifying} focusing on motion estimation, depth estimation and optical flow estimation.
However, such a framework is only evaluated on a desktop CPU and no quantitative results are provided.
Overall, all of these implementations are insufficient to fully unlock the potential advantages of event cameras for EMVS systems.

This motivates us to explore more efficient EMVS algorithm-hardware co-design approach for real-time target on low-power embedded platforms.
From comparative analysis, we observed that the event-based space-sweep procedures in EMVS have significant advantages including relatively high parallelism, low data dependency and low computational redundancy.
These advantages make it very suitable for customized hardware acceleration, which is adopted as the basic framework for our algorithm-hardware co-design and optimizations.

In this paper, \texttt{Eventor} is proposed as an FPGA/ARM heterogeneous accelerator for EMVS systems. The most time-consuming tasks of event back-projection and volumetric ray-counting are performed on FPGA. The main contributions are listed below:

\begin{itemize}[topsep=4pt,leftmargin=12pt]
\item[$\checkmark$] A novel efficient EMVS accelerator, \texttt{Eventor}, is proposed for real-time applications on embedded FPGA platform via algorithm-architecture co-design approaches.
\item[$\checkmark$] The involved EMVS algorithm is redesigned and customized in a hardware-friendly manner, which makes the accelerator much more efficient.
\item[$\checkmark$] Highly paralleled and fully pipelined architecture is designed and integrated with the heterogeneous execution model to improve the throughput and reduce the memory footprint.
\end{itemize}

The remainder of the paper is organized as follows. Section \ref{sec:system} demonstrates some comprehensive analysis of EMVS algorithm for potential optimization. Section \ref{sec:arch} illustrates the detailed architecture of the proposed \texttt{Eventor}. Evaluation results are provided in Section \ref{sec:exp}. Finally the conclusions are given in Section \ref{sec:con}.

%% file: 2-system.tex
\section{EMVS System}
\label{sec:system}

In this section, typical EMVS algorithm is analyzed for computational patterns evaluation and reformulated for hardware-friendly targeting.
Meanwhile, data quantization and compression strategies are further exploited to improve the computational efficiency.

\begin{figure}[t]
  \includegraphics[width=\linewidth]{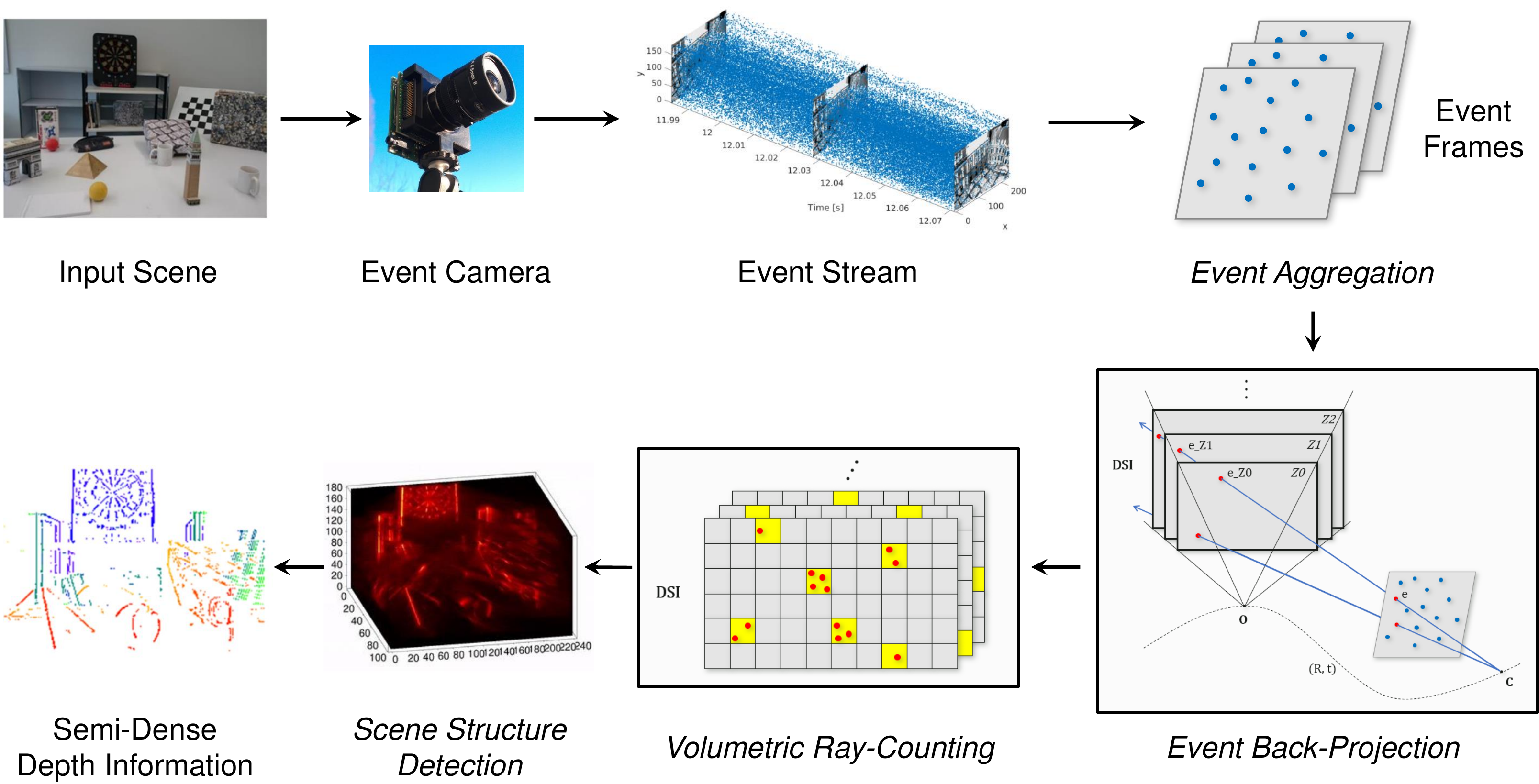}
  \caption{A typical EMVS workflow. This paper focuses on building the semi-dense depth information from the event streams.}
  \label{fig:emvs-workflow}
\end{figure}

\subsection{Algorithm Analysis}
\begin{figure}[t]
  \includegraphics[width=0.85\linewidth]{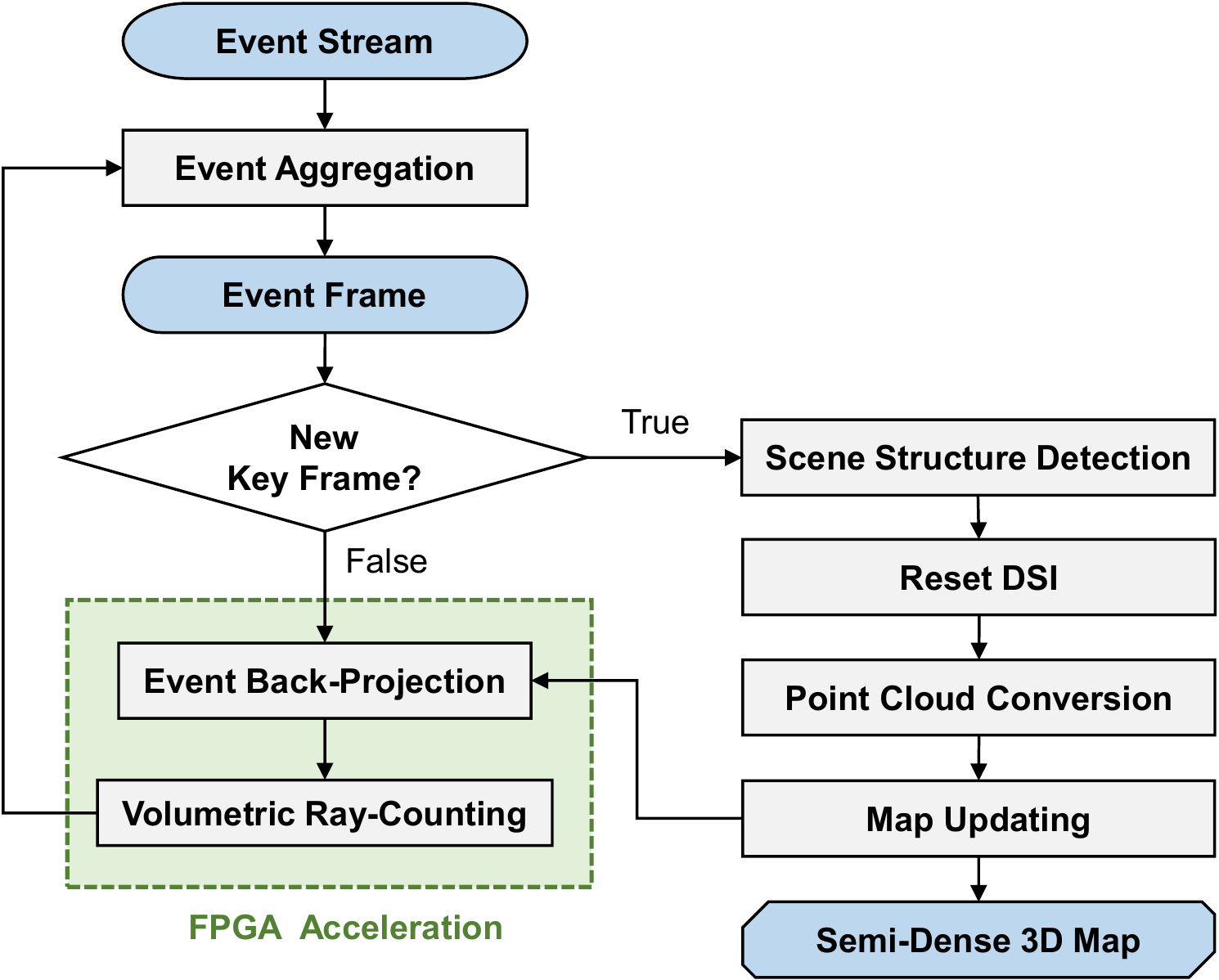}
  \caption{EMVS algorithm framework.}
  \label{fig:emvs-framework}
\end{figure}

EMVS algorithm aims to address the problem of estimating 3D structure from the event stream acquired by a moving event camera with a known trajectory \cite{rebecq2018emvs}.
A typical EMVS system is depicted in Fig.~\ref{fig:emvs-workflow}.
It mainly consists of four procedures: event \emph{aggregation} ($\mathcal{A}$), event back-\emph{projection} ($\mathcal{P}$), volumetric \emph{ray}-counting ($\mathcal{R}$) and scene structure \emph{detection} ($\mathcal{D}$).
The system receives the input event stream and corresponding camera trajectory, and reconstructs the semi-dense depth information of the viewing scene by event-based space-sweep method.
The complete workflow of EMVS algorithm is illustrated in Fig.~\ref{fig:emvs-framework}, and each stage is described as follows.


\emph{\textbf{Event Aggregation}}.
Specifically, when the logarithmic brightness at a certain pixel $\left(x_{k},y_{k}\right)$ reaches a threshold, event camera generates an event $e_{k}\doteq \left \langle x_{k},y_{k},t_{k},p_{k}\right \rangle$, where $x_{k}$ and $y_{k}$ is the corresponding pixel's coordinates of $k$-th event, $t_{k}$ is the timestamp of the triggered event and $p_{k}$ is the polarity of the brightness change.
\emph{Aggregation} (denoted as $\mathcal{A}$) divides the generated event stream to event frames (i.e. event packets) which will be processed together.

\emph{\textbf{Event Back-Projection}}.
Event back-\emph{projection} (denoted as $\mathcal{P}$) is the first stage of event-based space-sweep method. Each event in an event frame is back-projected to the viewing space according to the camera pose of the frame. 
Usually a ray density volume is created to record the distribution of back-projected rays.
A disparity space image (DSI) is interchangeably used to describe the discretized space volume and the scores stored in each voxel (i.e., the number of back-projected viewing rays passing through each voxel) \cite{rebecq2018emvs}.

The DSI is defined by dividing the viewing space to $N_{z}$ slices along the depth and discretizing each slice to $w \times h$ cuboid voxels, where $w$ and $h$ are the horizontal and vertical resolution of the event camera.
So the DSI size is $w \times h \times N_{z}$. Assuming the center of a voxel is $\mathcal{X}=\left(X,Y,Z\right)^{T}$, then back projecting events to the DSI can be discretized to the execution of mapping events to all the depth planes $\left \{ \mathbb{Z}_{i} \right \}_{i=1}^{N_{z}}$ located in the middle of the slices.

By creating a virtual camera located at a reference viewpoint, a DSI could be defined for its view recording.
The event back-projection is performed by two steps:
\ding{182}
Each event is firstly mapped from the current camera to the virtual camera via a canonical plane $\mathbb{Z}_{0}$ using homography matrix $\mathcal{H}_{\mathbb{Z}_{0}}$, which are denoted as $\mathcal{P}\left(\mathbb{Z}_{0}\right)$. The coordinates of events back-projected to $\mathbb{Z}_{0}$ are denoted as $\left \{x_{k}\left(\mathbb{Z}_{0}\right), y_{k}\left(\mathbb{Z}_{0}\right)\right \}$.
\ding{183}
The other depth planes $\mathbb{Z}_{i}$ could be obtained by mapping the points from $\mathbb{Z}_{0}$, which are denoted as $\mathcal{P}\left(\mathbb{Z}_{0} \leadsto \mathbb{Z}_{i}\right)$. The coordinates of events back-projected to $\mathbb{Z}_{i}$ are denoted as $\left \{x_{k}\left(\mathbb{Z}_{i}\right), y_{k}\left(\mathbb{Z}_{i}\right)\right \}$.

\emph{\textbf{Volumetric Ray-Counting}}.
After back-projecting events to DSI volume, the second stage of event-based space-sweep method is counting the number of back-projection rays that pass through each voxel (denoted as $\mathcal{R}$).
In the previous stage, the ray-voxel intersections are discretized to back projecting events to depth planes $\left \{ \mathbb{Z}_{i} \right \}_{i=1}^{N_{z}}$.
Then accumulating votes in the DSI can be done by voting DSI voxels at positions of $\left \{x_{k}\left(\mathbb{Z}_{i}\right), y_{k}\left(\mathbb{Z}_{i}\right), \mathbb{Z}_{i}\right \}$. 


\emph{\textbf{Key Frame Selection}}.
The EMVS algorithm selects several key reference views along the trajectory of the event camera and constructs local DSI.
After setting the original reference viewpoint, a new event frame could be only selected as a new key frame ($\mathcal{K}$) if the distance between the current event camera pose and the previous key reference view exceeds a threshold.
All of the events between two key frames will be utilized to estimate the local depth information.

\emph{\textbf{Scene Structure Detection}}.
Scene structure detection ($\mathcal{D}$) is the last stage of event-based space-sweep method.
A semi-dense depth map at the reference viewpoint is extracted from the DSI by determining whether a 3D point is present in each DSI voxel.
Based on the theory that the regions where multiple back-projection rays nearly intersect are likely to possess scene points, the algorithm determine 3D points by finding DSI voxels whose ray density scores are at local maximum of the ray density function.

\emph{\textbf{Merging Depth Information}}.
After getting the semi-dense depth map of the previous reference view at the scene structure detection procedure, the old local DSI is abandoned and a new local DSI is set in the viewing space of the new reference viewpoint.
Then the depth map is converted to a local point cloud and merged into the global point cloud ($\mathcal{M}$).
Hence, it includes three steps: reset DSI, point cloud conversion, and map updating.

\emph{\textbf{Computational Evaluation}}.
According to our observations, the most computational intensive and time-consuming tasks in the whole algorithm is \emph{event back-projection} ($\mathcal{P}$) and \emph{volumetric ray-counting} ($\mathcal{R}$).
When evaluating the EMVS algorithm on the DAVIS event camera dataset \cite{mueggler2017event}, the runtime of these two tasks accounts for over $80\%$ of total runtime.
To execute EMVS efficiently in real-time on a low-power embedded system, optimizations for these two tasks are obviously required, from both algorithm and hardware perspectives.
Hence, the procedures of $\mathcal{P}$ and $\mathcal{R}$ are accelerated by FPGA in our proposed \texttt{Eventor}.

\subsection{Hardware-Friendly Reformulation}
\label{subsec:reform}
\begin{figure}[t]
  \includegraphics[width=\linewidth]{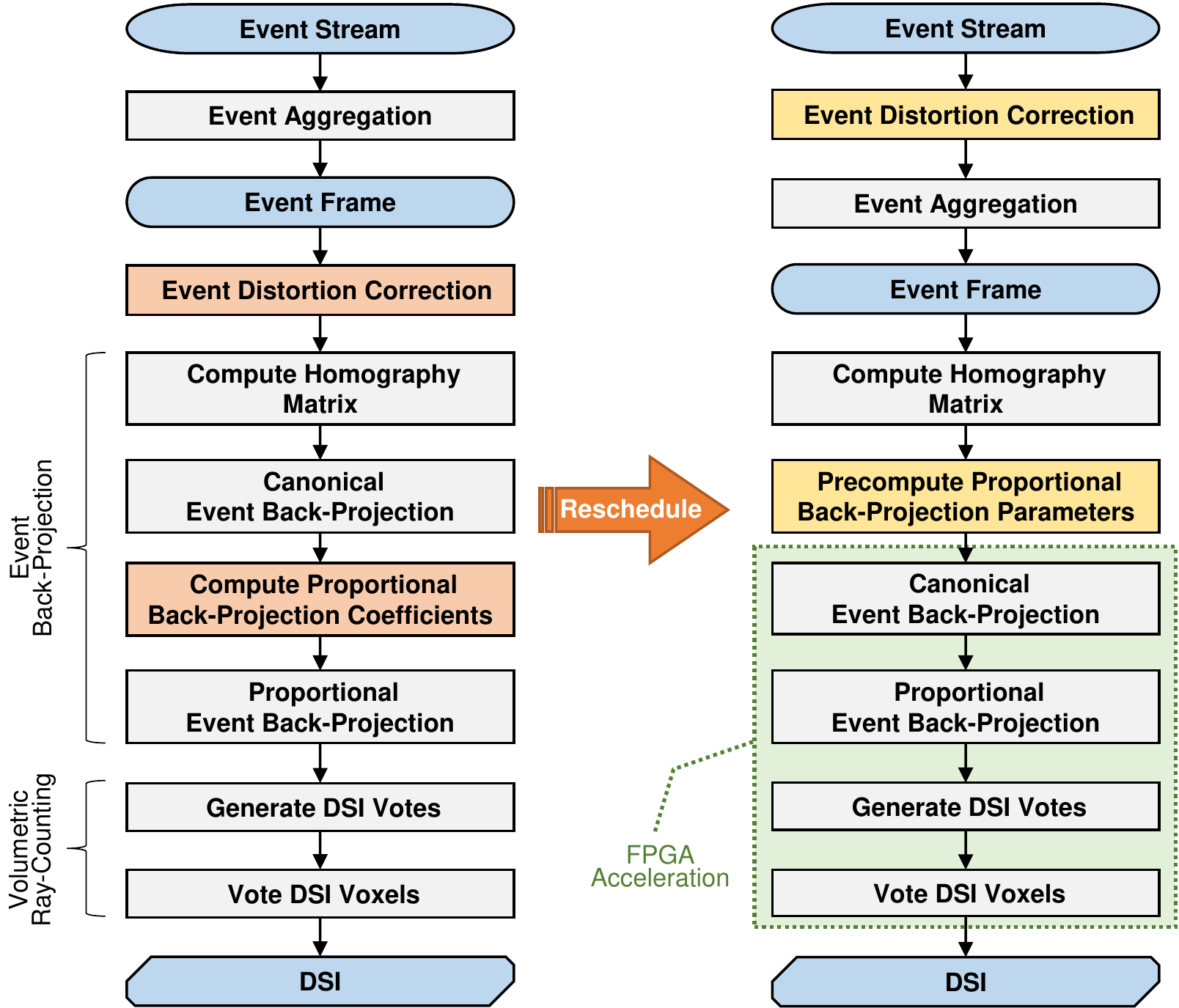}
  \caption{Details of original EMVS framework (left) and rescheduled for hardware-friendly optimization in our \texttt{Eventor} (right).}
  \label{fig:emvs-reschedule}
\end{figure}

Aiming to relieve the computational bottleneck ($\mathcal{P}$ and $\mathcal{R}$) of EMVS algorithm, an algorithm-hardware co-optimization approach is proposed where the original algorithm is rescheduled in a hardware-friendly manner as shown in Fig.~\ref{fig:emvs-reschedule}.
The event back-projection ($\mathcal{P}$) is divided into four sub-tasks:
\ding{202} \emph{Compute Homography Matrix} aims to compute the homography matrix $\mathcal{H}_{\mathbb{Z}_{0}}$,
\ding{203} \emph{Canonical Event Back-Projection} corresponds to $\mathcal{P}\left(\mathbb{Z}_{0}\right)$,
\ding{204} \emph{Compute Proportional Back-Projection Parameters} determines the parameters $\phi$ required in $\mathcal{P}\left(\mathbb{Z}_{0} \leadsto \mathbb{Z}_{i}\right)$,
\ding{205} \emph{Proportional Event Back-Projection} conducts the actual $\mathcal{P}\left(\mathbb{Z}_{0} \leadsto \mathbb{Z}_{i}\right)$. 
And the volumetric ray-counting ($\mathcal{R}$) is divided into two sub-tasks:
\emph{Generate DSI Votes} ($\mathcal{G}$) and \emph{Vote DSI Voxels} ($\mathcal{V}$).

\emph{\textbf{Workload Evaluation}}.
We further evaluate the computational workload of each sub-task for the above $\mathcal{P}$ and $\mathcal{R}$ procedures.
Among all the sub-tasks, \emph{Canonical Event Back-Projection} ($\mathcal{P}\left(\mathbb{Z}_{0}\right)$), \emph{Proportional Event Back-Projection} ($\mathcal{P}\left(\mathbb{Z}_{0} \leadsto \mathbb{Z}_{i}\right)$), \emph{Generate DSI Votes} ($\mathcal{G}$) and \emph{Vote DSI Voxels} ($\mathcal{V}$) will take up most of the runtime, because the required executions are proportional to the number of input events, while the \emph{Homography Matrix} ($\mathcal{H}_{\mathbb{Z}_{0}}$) and \emph{Proportional Back-Projection Parameters} ($\phi$) are only updated once when a new event frame is received.
Validation results on the DAVIS dataset show that the four sub-tasks above are responsible for over $90\%$ execution time of $\mathcal{P}$ and $\mathcal{R}$ procedures.

\emph{\textbf{Computation Parallelism Analysis}}.
The above $\mathcal{P}$ and $\mathcal{R}$ procedures could be found with high parallel availability.
According to the mechanism of event-based space-sweep method, there are mainly three types of parallelism in workloads:

\begin{itemize}[topsep=3pt,leftmargin=12pt]
\item[$\medtriangleright$] \emph{Operator-Level Parallelism}.
For the involved matrix and vector calculations in the procedure of $\mathcal{P}$, multiple arithmetic logic units (ALUs) could be deployed for fine-grained parallelism.
\item[$\medtriangleright$] \emph{Event-Level Parallelism}.
The procedure $\mathcal{P}$ requires to back-project each input event to the viewing space separately and extract scene structure from the ray density volume, which does not require simultaneous event observations or event matching.
Hence, different events can be processed in parallel and the computation stages involved can be fully pipelined.
\item[$\medtriangleright$] \emph{DSI-Level Parallelism}.
Due to the discretized structure of DSI and depth planes $\left \{ \mathbb{Z}_{i} \right \}_{i=1}^{N_{z}}$, the procedure $\mathcal{P}$ for different depth planes can be executed in parallel, so can voting for different DSI voxels.
\end{itemize}

\begin{figure}[t]
     \centering
     \begin{subfigure}{0.495\linewidth}
         \centering
         \includegraphics[width=\linewidth]{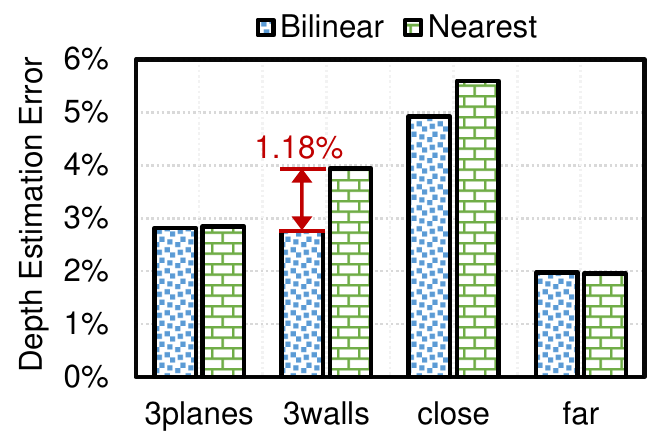}
         \caption{Different voting approaches.}
         \label{fig:error-compare:vote-error}
     \end{subfigure}
     \begin{subfigure}{0.495\linewidth}
         \centering
         \includegraphics[width=\linewidth]{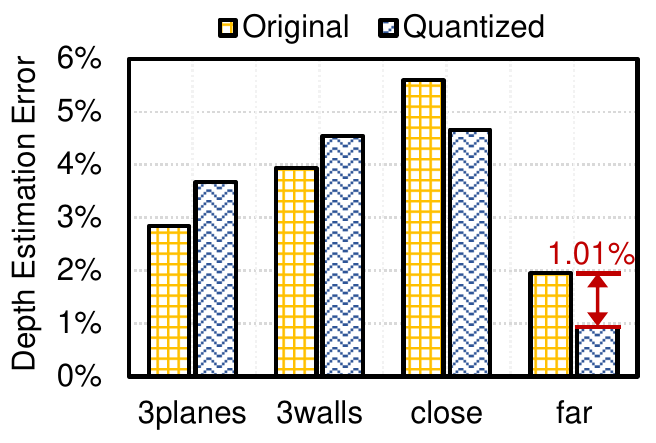}
         \caption{w/ or w/o quantization.}
         \label{fig:error-compare:quant-error}
     \end{subfigure}
\caption{Depth estimation error (AbsRel) comparison between different approaches evaluated on different datasets. The maximum AbsRel difference between Nearest Voting and original Bilinear Voting is about $1.18\%$. The maximum AbsRel difference before and after quantization is about $1.01\%$.}
\label{fig:error-compare}
\end{figure}

\emph{\textbf{Dataflow Reformulation}}.
According to the evaluation and analysis above, there are four tasks accelerated on FPGA: $\mathcal{P}\left(\mathbb{Z}_{0}\right)$, $\mathcal{P}\left(\mathbb{Z}_{0} \leadsto \mathbb{Z}_{i}\right)$, $\mathcal{G}$ and $\mathcal{V}$.
The high parallelism makes accelerating these tasks on FPGA rewarding.
However, the dataflow of the original EMVS framework shown in Fig.~\ref{fig:emvs-reschedule} (left) is not hardware-friendly enough.
Rescheduling the original algorithm to a streaming and hardware-friendly manner is proven to be an effective strategy in previous software-hardware co-optimization designs for traditional visual SLAM, such as the ORB-SLAM accelerator in \cite{liu2019eslam}.
Therefore, we perform reformulation to the EMVS algorithm for sufficient acceleration on heterogeneous systems.
As illustrated in Fig.~\ref{fig:emvs-reschedule} (right), the reformulation is mainly performed in the aspects of \emph{Rescheduling} and \emph{Approximate Computing}:

$\RHD$ \emph{\underline{Rescheduling}} includes the stages of \emph{Event Distortion Correction} and \emph{Compute Proportional Back-Projection Coefficients}.
\ding{182}
\emph{Event Distortion Correction} execution is originally performed after the events aggregated to a whole frame.
We set this stage before \emph{Event Aggregation} so that the correction is executed for each event in a streaming manner.
Streaming corrections could improve memory access efficiency during the aggregation stage.
\ding{183}
\emph{Proportional Back-Projection Coefficients} $\phi$ is pre-computed before performing $\mathcal{P}\left(\mathbb{Z}_{0}\right)$.
With the pre-computed $\phi$, the subsequent stages $\mathcal{P}\left(\mathbb{Z}_{0}\right)$, $\mathcal{P}\left(\mathbb{Z}_{0} \leadsto \mathbb{Z}_{i}\right)$, $\mathcal{G}$ and $\mathcal{V}$ could be efficiently accelerated on FPGA in parallel and fully pipelined. Meanwhile, the originally required data transfer of $\phi$ could be significantly reduced.

$\RHD$ \emph{\underline{Approximate Computing}} is adopted to improve the execution efficiency of procedure $\mathcal{R}$.
\ding{182} A standard DSI voting approach is named \emph{bilinear voting}, which is similar to bilinear interpolation.
Bilinear voting adopts a point $\left \{x_{k}\left(\mathbb{Z}_{i}\right), y_{k}\left(\mathbb{Z}_{i}\right), \mathbb{Z}_{i}\right \}$ to vote for the corresponding four nearest voxels on depth plane $\mathbb{Z}_{i}$ by splitting its contribution according to the distance between this point to each voxel.
\ding{183} Another approximate approach is called \emph{nearest voting}, which simply adopts each point to vote for its nearest neighboring voxels.
Nearest voting approach is less accurate than bilinear voting.
However, the computation complexity and memory access characteristics of nearest voting are much more hardware-friendly than bilinear voting.
The depth estimation accuracy comparison between Bilinear Voting and Nearest Voting is illustrated in Fig.~\ref{fig:error-compare:vote-error} by absolute relative error (AbsRel) across different datasets.
Fig.~\ref{fig:error-compare:vote-error} shows that the accuracy loss is acceptable when adopting nearest voting.
Considering the requirement of hardware-friendly manner, nearest voting is exploited in our dataflow.

\subsection{Hybrid Data Quantization}

Since most data involved in EMVS dataflow are represented by long floating-point format, we consider converting them as short fixed-point representations to reduce the memory footprint and data transferring bandwidth requirements.
Linear quantization method is utilized both for event coordinates and related parameters during the procedure of $\mathcal{P}$ and $\mathcal{R}$.
Detailed quantization strategies are illustrated in Table~\ref{tab:quantization}.

\begin{table}[t]
\caption{Detailed quantization strategies for procedure $\mathcal{P}$ and $\mathcal{R}$. Original floating-point data are quantized by fix-point data.}
\label{tab:quantization}
\small
\begin{tabular}{|c|c|c|c|}
\hline
Quantized Data Type & Total \#bit & \#bit of Integer & \#bit of Decimal  \\ \hline
($x_k$, $y_k$) & $16$ & $9$  & $7$  \\ \hline
$\left \{x_{k}\left(\mathbb{Z}_{0}\right), y_{k}\left(\mathbb{Z}_{0}\right)\right \}$ & $16$ & $9$ & $7$ \\ \hline
$\left \{x_{k}\left(\mathbb{Z}_{i}\right), y_{k}\left(\mathbb{Z}_{i}\right)\right \}$  & $8$ & $8$ & $0$  \\ \hline
$\mathcal{H}_{\mathbb{Z}_{0}}$ & $32$ & $11$ & $21$ \\ \hline
$\phi$   & $32$ & $11$ & $21$ \\ \hline
DSI Scores & $16$ & $16$ & $0$  \\ \hline
\end{tabular}
\end{table}

\emph{\textbf{Event Coordinates Quantization}}.
For event coordinates, we adopt a hybrid quantization strategy. 
Considering the byte-aligned bit width limitation and the $32$-bit data bus width between DRAM and FPGA, we utilize $16$-bit data to store the coordinates of the original input events ($x_k$, $y_k$).
In this way, the coordinates of an event are quantized as a pair of $16$-bit data and concatenated to a $32$-bit data to be saved in memory. 
For events generated by DAVIS event camera with resolution of $240 \times 180$, $9$-bit is enough for integer part of fixed-point coordinates, and remaining $7$-bit is exploited for decimal part.
Coordinates of $\left \{x_{k}\left(\mathbb{Z}_{0}\right), y_{k}\left(\mathbb{Z}_{0}\right)\right \}$ are quantized by using the same strategy. 
As for coordinates of $\left \{x_{k}\left(\mathbb{Z}_{i}\right), y_{k}\left(\mathbb{Z}_{i}\right)\right \}$, due to the mechanism of nearest voting method adopted in procedure $\mathcal{R}$, finding the nearest voxel to the projected point could be done by rounding the precise floating coordinates to integers.
Therefore, their coordinates can be quantized as $8$-bit integers.

\emph{\textbf{Parameters Quantization}}.
Since the homography matrix $\mathcal{H}_{\mathbb{Z}_{0}}$ and pre-computed parameters $\phi$ are usually invoked repeatedly during the procedures, their precision settings will have larger impact on the whole algorithm.
On the other hand, the required memory of these parameters are essentially much less than event coordinates and DSI scores.
As an appropriate strategy, they are quantized as $32$-bit data with $11$-bit integer part and $21$-bit decimal part.
As our observations, the sufficient integer bit width avoids data overflow, and continuing to increase the decimal bit width will not bring significant improvement to the depth estimation accuracy.

\emph{\textbf{DSI Scores Quantization}}.
For the scores stored in DSI voxels, they are quantized from $32$-bit float to $16$-bit integer.
Benefiting from nearest voting method, the increments (i.e. votes) of the scores are integer so that no decimal part is required.
Since the entire DSI structure are usually required to be stored in memory, such a quantization strategy can significantly reduce the memory footprint.

In summary, our hybrid data quantization strategy can save up to $50\%$ of the memory requirement and data transferring bandwidth.
Meanwhile, the depth estimation errors resulted from quantization are also evaluated across different datasets and illustrated in Fig.~\ref{fig:error-compare:quant-error}.
Evaluation results indicate that the accuracy of our quantized framework is comparable to the original full-precision framework.

%% file: 3-architecture.tex
\section{\texttt{Eventor} Architecture}
\label{sec:arch}

\begin{figure}[t]
  \includegraphics[width=\linewidth]{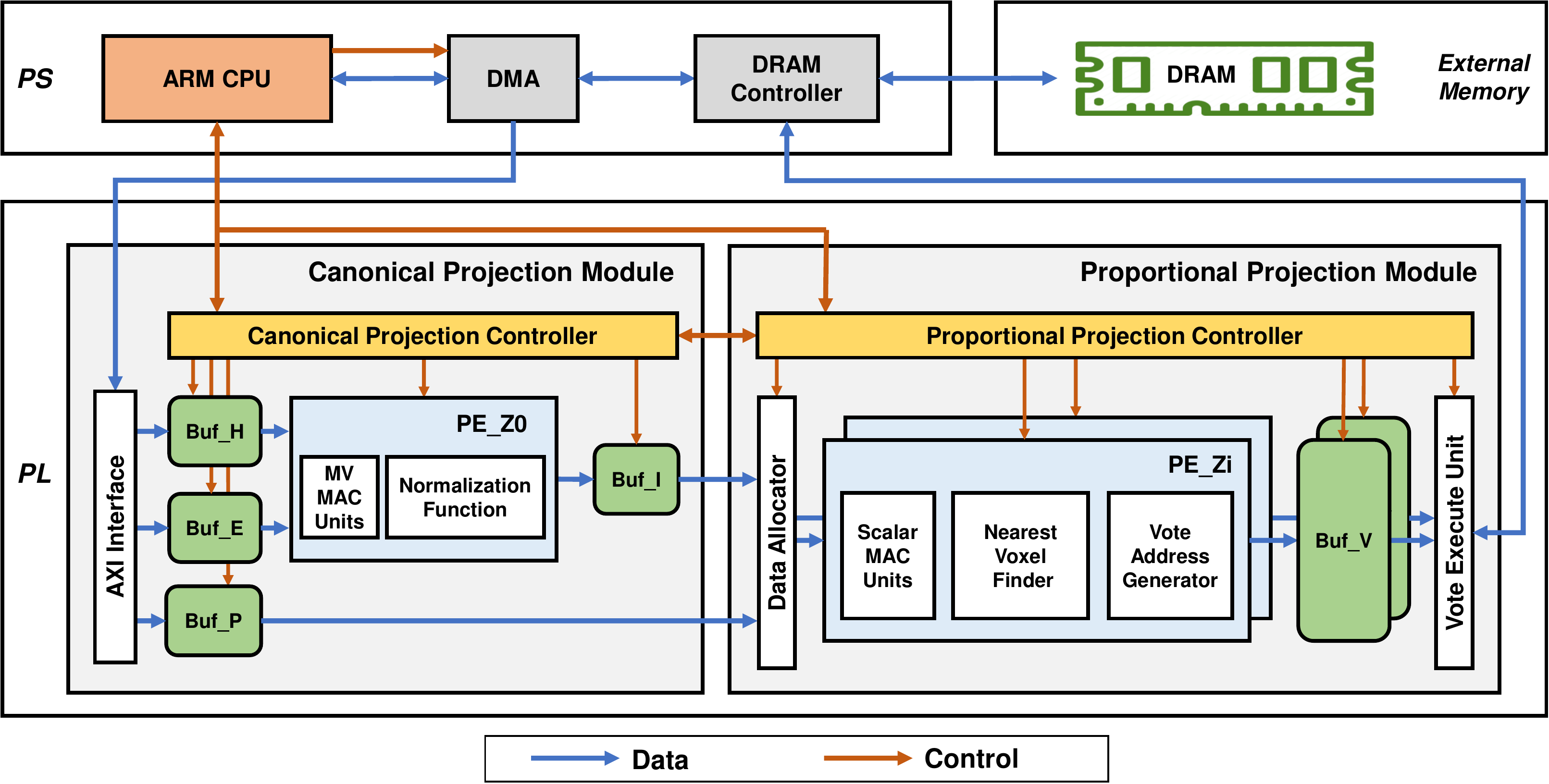}
  \caption{Overall hardware architecture of our proposed \texttt{Eventor}.}
  \label{fig:eventor-architecture}
\end{figure}

Base on the reformulated dataflow, overall hardware architecture of \texttt{Eventor} is designed on Zynq FPGA platform as shown in Fig.~\ref{fig:eventor-architecture}.
\texttt{Eventor} is partially implemented with programmable logic (PL) of FPGA and hosted by an ARM CPU as the processing system (PS).
\emph{Canonical Projection Module} and \emph{Proportional Projection Module} are exploited to compute $\mathcal{P}\left(\mathbb{Z}_{0}\right)$, $\mathcal{P}\left(\mathbb{Z}_{0} \leadsto \mathbb{Z}_{i}\right)$ and $\mathcal{R}$.
For processing each input event frame, ARM configures DMA to transfer input event coordinates $\left(x_{k},y_{k}\right)$ and parameters to input buffers.
Then ARM sends instructions to start the computational modules.
Overall, \texttt{Eventor} receives the input event frames streaming and updates the DSI data stored in DRAM.

\subsection{Canonical Projection Module}

Canonical Projection Module aims to compute $\mathcal{P}\left(\mathbb{Z}_{0}\right)$.
It receives the input event frames, $\mathcal{H}_{\mathbb{Z}_{0}}$, and outputs $\left \{x_{k}\left(\mathbb{Z}_{0}\right), y_{k}\left(\mathbb{Z}_{0}\right)\right \}$.
It also temporarily stores the proportional back-projection parameters and provides them together with intermediate event coordinates.

\textbf{AXI Interface} supports DMA to transfer input data and parameters via AXI bus. Quantized $16$-bit coordinates $\left(x_{k},y_{k}\right)$ are concatenated as 32-bit data which are transferred via AXI bus and stored in buffer.

\textbf{Buffers} in Canonical Projection Module include:
\ding{182}
\texttt{Buf\_H} for storing $\mathcal{H}_{\mathbb{Z}_{0}}$,
\ding{183}
Event Buffer \texttt{Buf\_E} for storing input event coordinates $\left(x_{k},y_{k}\right)$,
\ding{184}
Proportional Back-Projection Parameter Buffer \texttt{Buf\_P} for storing parameters $\phi$ required in $\mathcal{P}\left(\mathbb{Z}_{0} \leadsto \mathbb{Z}_{i}\right)$,
\ding{185}
Intermediate Buffer \texttt{Buf\_I} for storing $\left \{x_{k}\left(\mathbb{Z}_{0}\right), y_{k}\left(\mathbb{Z}_{0}\right)\right \}$.
Among them, \texttt{Buf\_H} is composed of registers since only one $3\times3$ homography matrix is required for each input event frame.
And the others are built with on-chip BRAM.
All of these buffers (including the Vote Buffer \texttt{Buf\_V} illustrated in Subsection \ref{sebsec:ppm} are realized by the manner of \emph{double-buffering}.
Many dataflow-driven accelerator designs have adopted this strategy to guarantee continuous loading and output streaming \cite{fu2018scalable}.
In this way, the transferring and processing of streaming data can be executed simultaneously, thus avoiding pipeline halt due to waiting for input data.


\textbf{PE\_Z0} is the processing element (PE) deployed in Canonical Projection Module for computing $\mathcal{P}\left(\mathbb{Z}_{0}\right)$.
It is equipped with a set of matrix-vector multiply-accumulate (MV MAC) units and a normalization function unit. $\mathcal{P}\left(\mathbb{Z}_{0}\right)$ is accelerated by multiple ALUs deployed in \texttt{PE\_Z0}, which are fully pipelined.
\texttt{PE\_Z0} loads $\mathcal{H}_{\mathbb{Z}_{0}}$ from \texttt{Buf\_H}, then receives streaming $\left(x_{k},y_{k}\right)$ from \texttt{Buf\_E} and outputs $\left \{x_{k}\left(\mathbb{Z}_{0}\right), y_{k}\left(\mathbb{Z}_{0}\right)\right \}$ to \texttt{Buf\_I}.
Since the workload of $\mathcal{P}\left(\mathbb{Z}_{0}\right)$ is less than $\mathcal{P}\left(\mathbb{Z}_{0} \leadsto \mathbb{Z}_{i}\right)$ and $\mathcal{R}$, only one \texttt{PE\_Z0} is deployed.
Besides, the latency of computing $\mathcal{P}\left(\mathbb{Z}_{0}\right)$ is not the critical path for normal frames in the pipelined workflow which will be demonstrated in Subsection \ref{sebsec:workflow}.

\textbf{Controller} in Canonical Projection Module mainly receives the starting instructions and configurations, then initializes \texttt{PE\_Z0} and buffers.
The Canonical Projection Controller is built as a finite-state machine (FSM), which has a specially designed synchronization state to synchronize the double-buffering state of \texttt{Buf\_E} together with the Proportional Projection Controller.
This synchronization mechanism ensures two modules to work in a pipelined mode.

\subsection{Proportional Projection Module}
\label{sebsec:ppm}

Proportional Projection Module is responsible for $\mathcal{P}\left(\mathbb{Z}_{0} \leadsto \mathbb{Z}_{i}\right)$ and $\mathcal{R}$.
It receives $\left \{x_{k}\left(\mathbb{Z}_{0}\right), y_{k}\left(\mathbb{Z}_{0}\right)\right \}$ and $\phi$ from Canonical Projection Module, and updates the DSI voxels scores.

\textbf{PE\_Zi:} Canonical Projection Module has multiple \texttt{PE\_Zi} to execute $\mathcal{P}\left(\mathbb{Z}_{0} \leadsto \mathbb{Z}_{i}\right)$ and $\mathcal{G}$.
\texttt{PE\_Zi} receives $\left \{x_{k}\left(\mathbb{Z}_{0}\right), y_{k}\left(\mathbb{Z}_{0}\right)\right \}$ and $\phi$ from Data Allocator, and generates the addresses of DSI voxels which are required for \texttt{Buf\_V}.
\texttt{PE\_Zi} includes: Scalar MAC Units, Nearest Voxel Finder and Vote Address Generator.
Scalar MAC Units execute $\mathcal{P}\left(\mathbb{Z}_{0} \leadsto \mathbb{Z}_{i}\right)$.
Nearest Voxel Finder computes the nearest DSI voxel to $\left \{x_{k}\left(\mathbb{Z}_{i}\right), y_{k}\left(\mathbb{Z}_{i}\right), \mathbb{Z}_{i}\right \}$ and conducts projection missing judgement.
Vote Address Generator generates the vote addresses, which are directly utilized for updating DSI scores.
Usually different PEs (multiple \texttt{PE\_Zi}) could share a same event input and operate simultaneously in parallel for different depth planes.

\textbf{Data Allocator} fetches input data and parameters required by \texttt{PE\_Zi} and allocates them to PEs.
Different PEs need different parameters while sharing a same event input.
The dataflow between \texttt{Buf\_I} and \texttt{PE\_Zi} is managed by this allocator.

\textbf{Vote Execute Unit} exploits the DSI vote addresses stored in \texttt{Buf\_V} to vote the corresponding voxels.
It is equipped with two AXI-HP ports and data transfer logic to directly access the DRAM via the DRAM controller, no need for ARM intervention. The old scores stored in DSI voxels are fetched from DRAM, added by a vote value (typically $1$) and wrote back to DRAM.

\begin{figure}[t]
  \includegraphics[width=\linewidth]{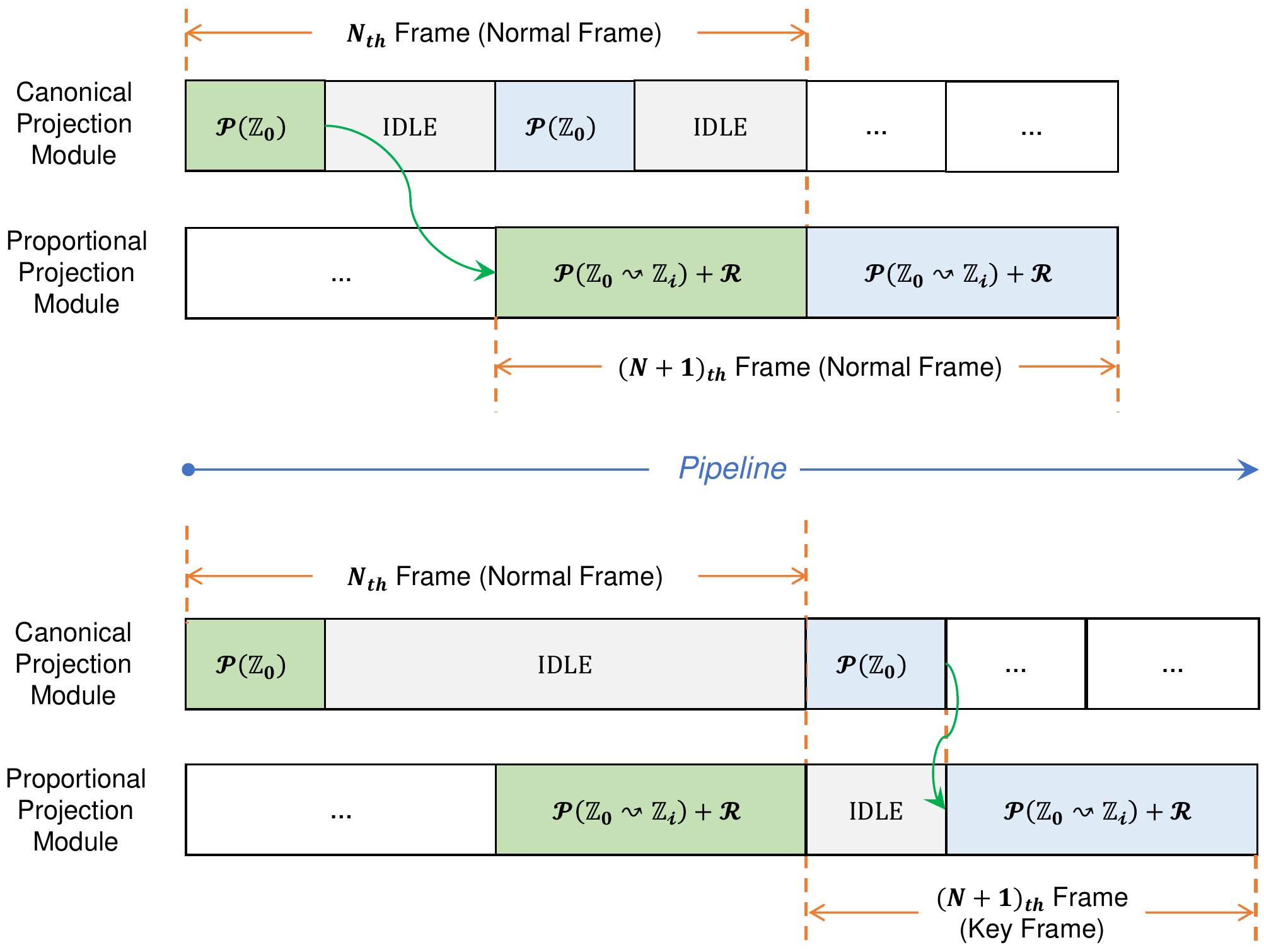}
  \caption{The pipelined workflow of normal event frame (upper) and key event frame (lower).} 
  \label{fig:pipeline}
\end{figure}

\subsection{Accelerator Workflow}
\label{sebsec:workflow}

The overall execution model of \texttt{Eventor} is shown in Figure \ref{fig:pipeline}.
Canonical Projection Module and the Proportional Projection Module work in a pipelined order while \texttt{Eventor} receives the streaming input event frames. 

For normal event frames, the two modules work simultaneously.
Canonical Projection Module starts working as soon as \texttt{Buf\_I} is ready for new input so that Proportional Projection Module can operate continuously.
In this way, the actual execution time for each frame is equal to the sum of the execution time of $\mathcal{P}\left(\mathbb{Z}_{0} \leadsto \mathbb{Z}_{i}\right)$ and $\mathcal{R}$, and the execution time of $\mathcal{P}\left(\mathbb{Z}_{0} \right)$ is overlapped.

Things are different when a new key event frame is selected.
Because a new key frame means a new reference view, the DSI will be reset and the following events will be back projected and vote for the new DSI.
So the Canonical Projection Module will wait until the Proportional Projection Module finishes processing the previous event frame, then start processing the key event frame if it is fired up.
The Proportional Projection Module then starts to work once receiving $\left \{x_{k}\left(\mathbb{Z}_{0}\right), y_{k}\left(\mathbb{Z}_{0}\right)\right \}$.
Therefore, the execution time for a key frame is equal to the sum of the execution time of $\mathcal{P}\left(\mathbb{Z}_{0} \right)$, $\mathcal{P}\left(\mathbb{Z}_{0} \leadsto \mathbb{Z}_{i}\right)$ and $\mathcal{R}$.

\subsection{Parallelization Mechanism}
According to the computation parallelism analysis carried out in Section \ref{subsec:reform}, three levels of parallelism are involved: operator-level, event-level and DSI-level.
\texttt{Eventor} aims to fully utilize these parallelism.
For operator-level parallelism, we deploy multiple ALUs in \texttt{PE\_Z0} to accelerate matrix and vector calculation.
For event-level parallelism, the workflow and datapath of \texttt{Eventor} is designed as a fully-pipelined scheme to process events without data dependency.
For DSI-level parallelism, multiple \texttt{PE\_Zi} are implemented inside the Proportional Projection Module to back-project an event to multiple depth planes and generate vote addresses simultaneously.
Benefiting from the exploration of parallelism, our \texttt{Eventor} is able to achieve a relatively high event processing rate.

%% file: 4-experiment.tex
\section{Experimental Results}
\label{sec:exp}

This section first introduces our experimental setup.
Then, we evaluate the effectiveness of our hardware-friendly dataflow reformulation and the proposed \texttt{Eventor} accelerator.

\subsection{Experimental Setup}
\textbf{Hardware Implementation:}
The \texttt{Eventor} is implemented and evaluated on Xilinx Zynq XC7Z020 SoC \cite{zynq}.
Its PL is with $4.9$ Mb BRAM as on-chip memory and 1 GB, 32-bit DDR3 DRAM as external memory. The clock frequency of \texttt{Eventor} is 130 MHz, and the DDR clock is $533$ MHz.
The prototype of \texttt{Eventor} is equipped with two \texttt{PE\_Zi} and corresponding \texttt{Buf\_I} in Proportional Projection Module.
The computation modules on PL are implemented by hand-optimised RTL.
The resources utilization of \texttt{Eventor} are shown in Table~\ref{tab:utilization}. It can be seen that \texttt{Eventor} uses quite few resources.

\begin{table}[t]
\caption{The FPGA resources utilization of \texttt{Eventor}.}
\label{tab:utilization}
\small
\begin{tabular}{|c|c|c|c|}
\hline
 & \# LUT & \# FF & BRAM  \\ \hline
Utilization & $ 17538$ ($32.97\%$) & $22830$ ($21.46\%$) & $64$ KB ($11.43\%$)  \\ \hline
\end{tabular}
\end{table}

\textbf{Dataset:} The reformulated EMVS framework and \texttt{Eventor} are evaluated on DAVIS event camera dataset and simulator \cite{mueggler2017event}.
It contains event streams captured with a DAVIS event camera in a variety of simulated and real environments, along with ground-truth camera trajectories.
The resolution of a DAVIS event camera is $240\times180$.
Four different sequences are used for evaluation: \emph{simulation\_3planes} and \emph{simulation\_3walls} are simulated sequences, \emph{slider\_close} and \emph{slider\_far} are captured in real scene.



\begin{figure}[t]
     \centering
     \begin{subfigure}{0.495\linewidth}
         \centering
         \includegraphics[width=\linewidth]{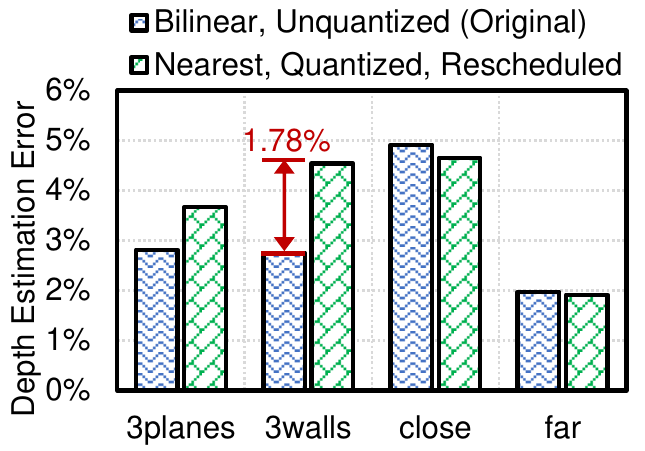}
         \caption{The depth estimation error (AbsREL) of our reformulated hardware-friendly EMVS when compared with original EMVS.}
         \label{fig:total-error:error}
     \end{subfigure}
     \hfill
     \begin{subfigure}{0.43\linewidth}
         \centering
         \includegraphics[width=\linewidth]{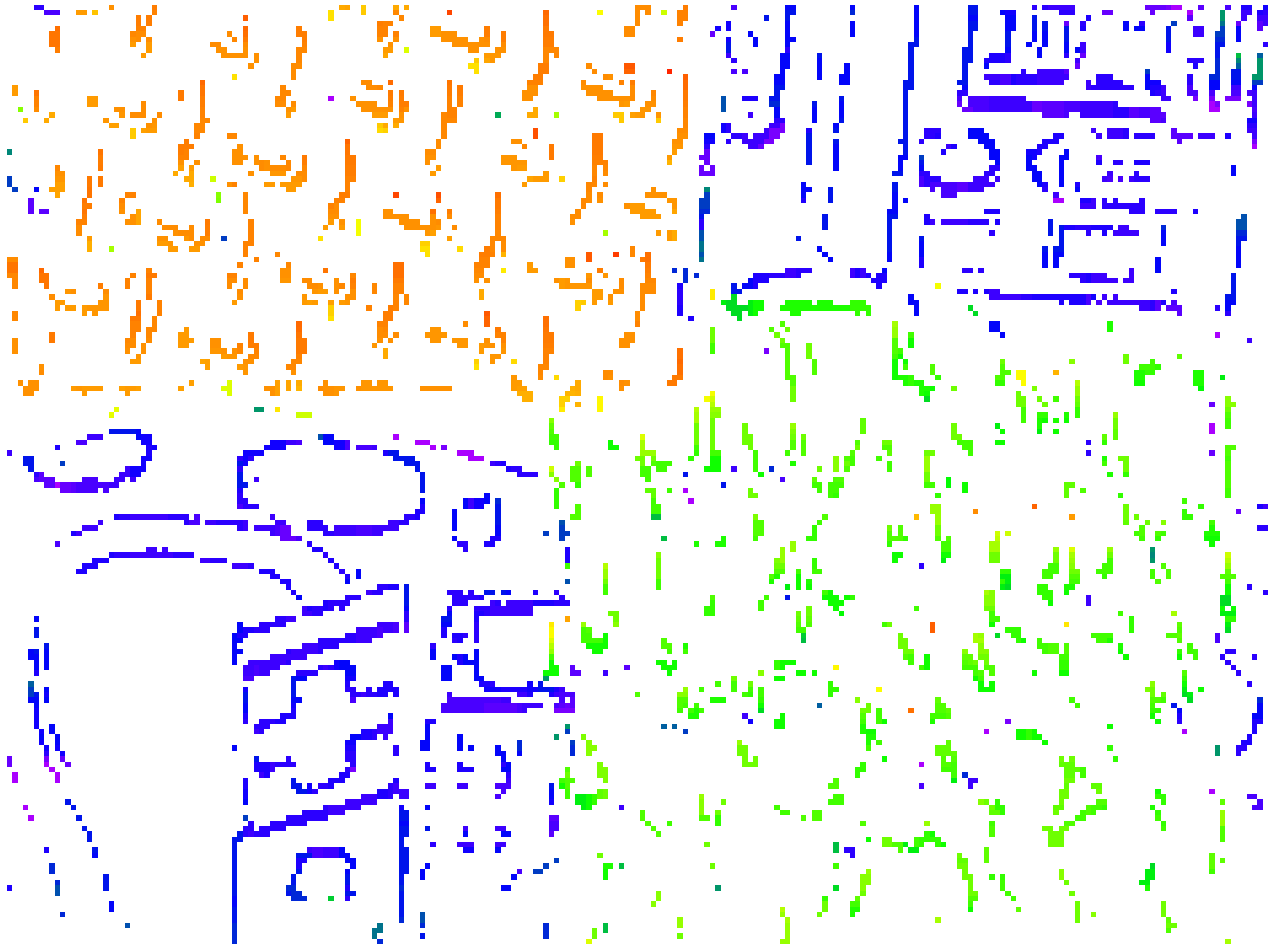}
         \caption{A sample demonstration of reconstructed scene structure from the sequence of \emph{simulation\_3planes}.}
         \label{fig:total-error:scene}
     \end{subfigure}
\caption{Accuracy of depth estimation comparison and reconstructed scene structure demonstration.}
\label{fig:total-error}
\end{figure}

\subsection{Accuracy Analysis}

The accuracy of EMVS is measured by depth estimation error (absolute relative error, AbsREL), which means the difference between the depth of reconstructed scene structure and the groundtruth.
Fig.~\ref{fig:total-error:error} shows the comparison of average depth estimation error between original EMVS and our reformulated framework. 
For \emph{simulation\_3planes} and \emph{simulation\_3walls}, the original EMVS has a better accuracy than our reformulated framework, but the maximum difference is less than $1.78\%$.
For \emph{slider\_close} and \emph{slider\_far}, our framework even has a better accuracy than the original EMVS.
Overall, the results indicate that the accuracy of our reformulated framework is comparable to original EMVS.
A sample reconstructed scene structure from the sequence of \emph{simulation\_3planes} is also demonstrated in Fig.~\ref{fig:total-error:scene} for 3D view.

\subsection{Performance Evaluation}
The performance of \texttt{Eventor} is compared with the EMVS run on Intel i5-7300HQ CPU \cite{intel}.
Comparison results of computation speed and power consumption are illustrated in Table~\ref{tab:performance}, including detailed runtime breakdown, average runtime per event frame and the event processing rate.
Each event frame consists of $1024$ events, which is determined according to the sensor's event rate and storage.


Compared with the Intel i5 CPU, the event processing rate of \texttt{Eventor} is slightly higher, without obvious advantage.
However, in terms of power consumption, \texttt{Eventor} shows great advantage over the Intel CPU. As shown in Table \ref{tab:performance}, the power consumption can be reduced by $24\times$.
\texttt{Eventor} is able to achieve significant energy reduction with no loss of performance.


\begin{table}[t]
\caption{Performance comparison between \texttt{Eventor} and original EMVS run on Intel i5 CPU.}
\label{tab:performance}
\small
\begin{tabular}{|cc|c|c|}
\hline
\multicolumn{2}{|c|}{} & Intel CPU & \texttt{Eventor}  \\ \hline
\multicolumn{1}{|c|}{\multirow{2}{*}{\begin{tabular}[c]{@{}c@{}}Runtime per Event Frame\\ ( $\mu$s / task )\end{tabular}}} & $\mathcal{P}\left(\mathbb{Z}_{0}\right)$ & 22.40 & 8.24 \\ \cline{2-4} 
\multicolumn{1}{|c|}{} & $\mathcal{P}\left(\mathbb{Z}_{0} \leadsto \mathbb{Z}_{i}\right)$ \& $\mathcal{R}$ & 559.55 & 551.58 \\ \hline
\multicolumn{1}{|c|}{\multirow{2}{*}{\begin{tabular}[c]{@{}c@{}}Runtime per Event Frame\\ ( $\mu$s / frame )\end{tabular}}} & Normal frame & 581.95 & 551.58 \\ \cline{2-4} 
\multicolumn{1}{|c|}{} & Key frame & 581.95 & 559.82 \\ \hline
\multicolumn{1}{|c|}{\multirow{2}{*}{\begin{tabular}[c]{@{}c@{}}Event Processing Rate\\ ( $10^6$ events / second)\end{tabular}}} & Normal frame & 1.76 & 1.86 \\ \cline{2-4} 
\multicolumn{1}{|c|}{} & Keyframe & 1.76 & 1.83 \\ \hline
\multicolumn{2}{|c|}{Power (W)} & 45 & 1.86 \\ \hline
\end{tabular}
\end{table}

%% file: 5-conclusions.tex
\section{Conclusions}
\label{sec:con}
In this paper, an efficient EMVS accelerator, \texttt{Eventor}, is proposed for real-time applications and evaluated on Zynq FPGA platform.
The EMVS algorithm is partly reformulated to a more hardware-friendly manner, and hybrid data quantization strategies are adopted to improve the computational efficiency.
Meanwhile, the most time-consuming stages, i.e., event back-projection and volumetric ray-counting are accelerated on FPGA with different parallelism.
Evaluation results show that \texttt{Eventor} could achieve $24\times$ improvement in energy efficiency compared with Intel i5 CPU.
The overall performance of \texttt{Eventor} could satisfy the requirements of real-time reconstruction on power-limited embedded platforms.